\documentclass[12pt]{article}
\usepackage{epsbox}
\begin{document}

\title{The gluonic excitation of  the three-quark system in SU(3) lattice QCD}

\author{Toru~T.~Takahashi
\thanks{Research Center for Nuclear Physics, Osaka University,
Mihogaoka 10-1, Ibaraki 567-0047, Japan}
and Hideo~Suganuma
\thanks{Faculty of Science, Tokyo Institute of Technology,
Ohokayama 2-12-1, Tokyo 152-8551, Japan}}

\maketitle

\begin{abstract}
We present the first study of the gluonic excitation in the three-quark (3Q) system 
in SU(3) lattice QCD with $\beta$=5.8 and $16^3 \times 32 $ at the quenched level. 
For the spatially-fixed 3Q system, we measure the gluonic excited-state potential,  
which corresponds to the flux-tube vibrational energy in the flux-tube picture. 
The lowest gluonic-excitation energy in the 3Q system is found to be about 1GeV in the hadronic scale. 
This large gluonic-excitation energy is expected to bring about the 
success of the simple quark model without gluonic modes.
\end{abstract}

The low-lying hadrons are well categorized with the simple 
quark model,  
which has only quark degrees of freedom.
The success of the quark model implies the absence of the gluonic excitation 
in the low-lying hadron spectra, which seems rather mysterious.
To understand the success of the quark model based on QCD, 
it is important to investigate the gluonic excitation mode.

Theoretically, the gluonic excitation is expected to appear as the vibrational 
mode of  the squeezed color-flux-tube, which is formed due to color confinement \cite{KS75,CI86}.
Experimentally, this type of the gluonic excitation is 
closely related to the hybrid mesons or the hybrid baryons, 
which consist of $q\bar qG$ or $qqqG$, respectively, in the valence picture.
In particular, the several states with the exotic quantum number such as $J^{PC}=0^{--},0^{+-},1^{-+},2^{+-},...$ 
cannot be constructed with the simple quark picture\cite{DGH92}, 
and therefore they are now investigated 
with much attention from both theoretical and experimental viewpoints \cite{CP02}.

For the Q-$\bar {\rm Q}$ system corresponding to mesons, 
there are several lattice QCD studies for both the ground-state 
and the excited-state potentials. 
The lattice QCD study indicates that the Q-$\bar{\rm Q}$ ground-state potential 
takes the form of $V_{\rm Q\bar Q}^{\rm g.s.}(r)=-A\frac1{r}+\sigma r$,
as the function of the inter-quark distance $r$,
with $A \simeq 0.27$ and 
$\sigma\simeq$0.89GeV/fm besides a irrelevant constant \cite{TMNS01,TSNM02}.
On the other hand, the lowest excited-state potential for the spatially-fixed
Q-$\bar {\rm Q}$ system can be fitted as 
$V_{\rm Q\bar Q}^{\rm e.s.}(r)=\sqrt{b_0+b_1r+b_2r^2}+c_0$ \cite{JKM98}.
Here, the quantum number of the first-excited state is 
$L=1$ for the magnitude of the projection of the total 
angular momentum of the gluon onto the Q-$\bar{\rm Q}$ axis \cite{JKM98}.
The excitation energy $V_{\rm Q\bar Q}^{\rm e.s.}-V_{\rm Q\bar Q}^{\rm g.s.}$ is in the order of 1GeV 
in the typical hadronic scale as 0.5$-$1.0fm, and seems rather large in comparison with 
the low-lying excitation energy of the quark origin.

For the 3Q system corresponding to baryons, the situation is much more complicated.
Even for the 3Q ground-state potential, the accurate measurement of the 3Q potential 
using lattice QCD is relatively difficult and has been  performed recently. 
The recent detailed lattice-QCD analysis indicates 
the Y-type flux-tube formation in the ground-state 3Q system~\cite{TMNS01,TSNM02,I02}.
The vibrational modes of the Y-type flux tube system are considered to be 
much more complicated and chaotic than those of the simple Q-$\bar{\rm Q}$ flux-tube, 
owing to the possible interference among the vibrational modes on the three-flux tubes 
connected at the physical junction, which is not spatially fixed. 

Note here that the magnitude of the excitation energy is closely related to the nature of the physical junction. 
For instance, if the physical junction behaves as a quasi-fixed edge of the three flux tubes, 
the vibrational energy in the 3Q system would be expressed as a simple sum of the three modes on 
the Q-$\bar {\rm Q}$ system, where the quark and antiquark behave as fixed edges.
If the physical junction behaves as a quasi-free edge of the flux tube, 
the vibrational energy in the 3Q system would be smaller than that in the Q-$\bar {\rm Q}$ 
system with the similar size.
(In general in the quantum physics, the vibrational modes on the string with the fixed edges becomes large.)

To begin with, we briefly review the lattice-QCD measurement of 
the 3Q ground-state potential 
from the 3Q Wilson loop using the smearing method~\cite{TMNS01,TSNM02}.
\begin{figure}[h]
\begin{center}
\epsfile{file=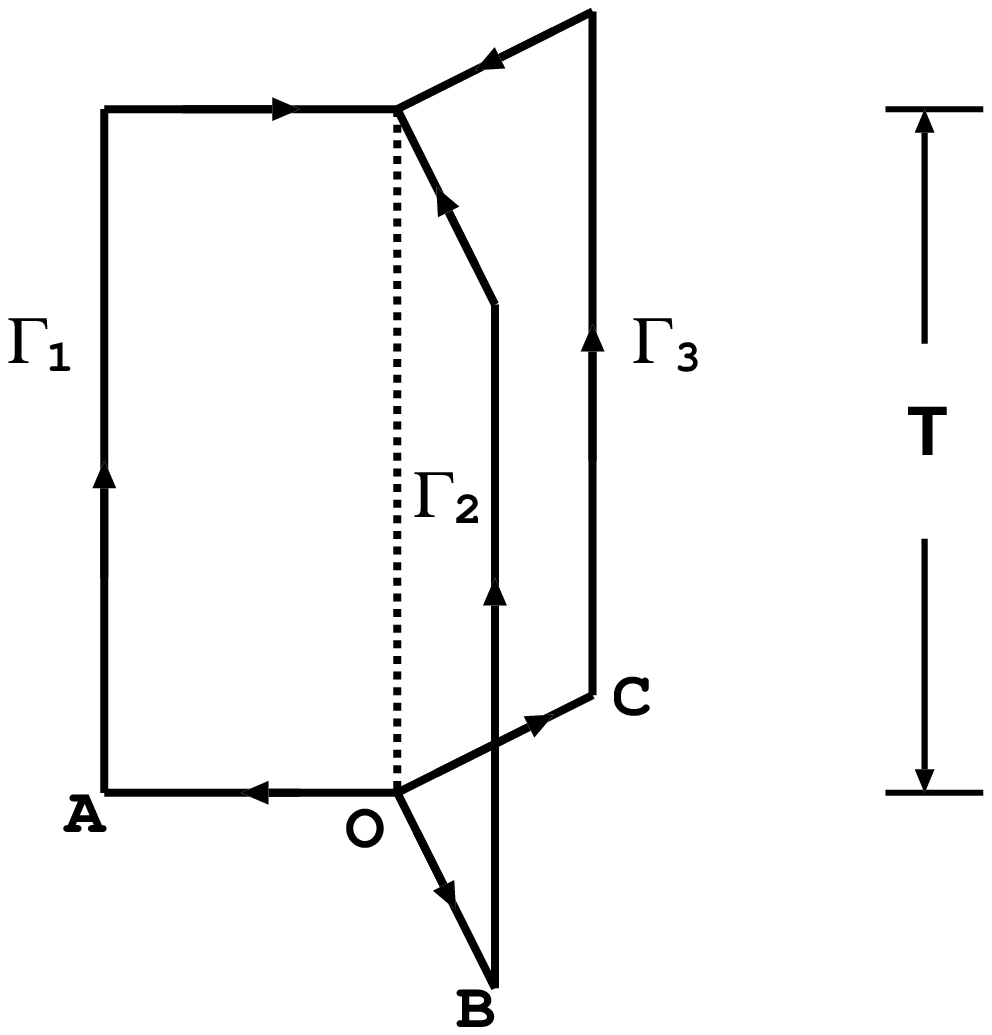,scale=0.35}
\end{center}
\caption{\label{3Qloop}
The 3Q Wilson loop $W_{\rm 3Q}$.
The gauge-invariant 3Q state is generated at $t=0$ and is annihilated at $t=T$. 
The three quarks are spatially fixed in ${\bf R}^3$ for $0 < t < T$.
}
\end{figure}
Similar to the Q-${\rm\bar Q}$ potential measured with the Wilson loop,
the 3Q potential can be measured with the 3Q Wilson loop $W_{\rm 3Q}$ 
defined as 
\begin{equation}
W_{\rm 3Q} \equiv \frac1{3!}\varepsilon_{abc}\varepsilon_{a'b'c'}
U_1^{aa'} U_2^{bb'} U_3^{cc'} 
\end{equation}
with $U_k \equiv {\rm P}\exp\{ig\int_{\Gamma_k}dx^\mu A_{\mu}(x)\}$ 
($k=1,2,3$).  
Here, P denotes the path-ordered product along the path denoted by 
${\Gamma_k}$ in Fig.~\ref{3Qloop}.
The ground-state potential $V_{\rm 3Q}^{\rm g.s.}$ can be obtained from 
the expectation value  $\langle W_{\rm 3Q}(T)\rangle$ of the 3Q Wilson loop 
in the large limit of  its temporal length $T$ 
as 
$V_{\rm 3Q}^{\rm g.s.}=
-{\lim}_{T\rightarrow 0}\frac{1}{T}\ln \langle W_{\rm 3Q}(T)\rangle$.
In the practical lattice calculation, 
one has to take enough large $T$ where the excited-state contributions drop, 
but $\langle W_{\rm 3Q}(T)\rangle$ itself exponentially decreases with $T$, 
which makes the accurate measurement difficult. 

Then, for the accurate measurement, 
we apply the smearing method, a useful standard technique to 
enhance the ground-state component in the 3Q operator $W_{\rm 3Q}$~\cite{TMNS01,TSNM02}.
The smearing is defined as the iterative replacement of the spatial link variables by
the ``fat'' link variables, and refines the 3Q operator 
to have a large overlap with the ground state of the 3Q system.
In fact, spatial link variables $U_i(s)$ are replaced
by the new link variables $\bar U_i(s)$ defined so as to maximize
\begin{eqnarray}
{\rm Re} \,\,{\rm tr}\Bigg\{ \bar U_i^{\dagger}(s) \Bigg[
\alpha U_i(s)+\sum_{j \ne i} \{
U_j(s)U_i(s+\hat j)U_j^\dagger (s+\hat i)
+U_j^\dagger (s-\hat j)U_i(s-\hat j)U_j(s+\hat i-\hat j)
\} \Bigg] \Bigg\},
\end{eqnarray}
where we adopt $\alpha=2.3$ for the smearing parameter. 
This replacement is iterated until 
the overlap of the 3Q operator with the ground state is maximized, i.e.,  
the contamination of the excited states is minimized.
For instance, as the iteration number $N_{\rm smr}$ of the smearing increases from $N_{\rm smr}$=0, 
the excited-state components in the $N_{\rm smr}\hbox{ -th}$ smeared operator 
gradually decrease, and finally the 22th smeared operator is almost ground-state 
saturated for $\alpha=2.3$ on the lattice with $\beta=5.8$ \cite{TSNM02}.

Using the suitable
smeared link-variable, we reconstruct 
the generalized 3Q Wilson loop $\langle W_{\rm 3Q}(T)\rangle$.
Then, one finds $V_{\rm 3Q}^{\rm g.s.} \simeq -\frac{1}{T}\ln \langle W_{\rm 3Q}(T)\rangle$ even at small $T$, 
and one can perform the accurate measurement of $V_{\rm 3Q}^{\rm g.s.}$.
From the recent detailed analysis of the static 3Q potential 
in Refs.~\cite{TMNS01,TSNM02},  
the ground-state 3Q potential 
$V_{\rm 3Q}^{\rm g.s.}$ 
is found to be expressed as
\begin{equation}
V_{\rm 3Q}^{\rm g.s.}=-A_{\rm 3Q}\sum_{i<j}\frac1{|{\bf r}_i-{\bf r}_j|}
+\sigma_{\rm 3Q} L_{\rm min}+C_{\rm 3Q}, 
\end{equation}
where $L_{\rm min}$ denotes the minimal value of 
the total length of color-flux-tubes linking the three quarks~\cite{KS75,CI86,BPV95}.
(The observation of the Y-type flux-tube profile is also reported in lattice QCD \cite{I02}.)

Next, we present the formalism to extract the excited-state potential.
For the simple notation, the ground state is regarded as the ``0-th excited state'' in this paper.
For the physical eigenstates of the QCD Hamiltonian $\hat H$ 
for the spatially-fixed 3Q system, we denote 
the $n$-th excited state by $|n \rangle$ ($n$=0,1,2,3,..).
Since the three quarks are spatially fixed in this case, 
the eigenvalue of $\hat H$ is expressed by the static 3Q potential as 
\begin{eqnarray}
\hat H|n\rangle=E_n|n\rangle=V_n|n\rangle, 
\end{eqnarray}
where 
$V_n$ denotes the $n$-th excited-state 3Q potential. 
We take the normalization condition as $\langle m|n \rangle=\delta_{mn}$.
Note that both $V_n$ and $|n \rangle \ (n=0,1,2,..)$ are 
universal physical quantities relating to the QCD Hamiltonian $\hat H$.

Suppose that $|\Phi_k \rangle \ (k=0,1,2,3,..)$ 
are arbitrary given independent 3Q operators 
for the spatially-fixed 3Q system. In general, each 3Q operator 
$|\Phi_k \rangle$ can be expressed with a linear combination 
of the physical eigenstates $|n\rangle \ (n=0,1,2,..)$ as 
\begin{eqnarray}
|\Phi_k \rangle =c_0^k|0\rangle+c_1^k|1\rangle+c_2^k|2\rangle+..
\end{eqnarray}
Here, the coefficients $c_n^k$ depend on the selection of the 3Q operators 
$|\Phi_k \rangle$, and hence they are not universal quantities.
(Unlike the Q-$\bar{\rm Q}$ system, there is no definite symmetry 
in the 3Q system, and hence we do not construct the operators which
carry the specific quantum number.)

The Euclidean time-evolution of the 3Q state $|\Phi(t)\rangle$ is expressed with the operator $e^{-\hat Ht}$, 
which corresponds to the transfer matrix in lattice QCD. 
The overlap $\langle \Phi_j(T)|\Phi_k(0)\rangle$ is given by 
the 3Q Wilson loop sandwiched by the initial state $\Phi_k$ at $t=0$ and the final state $\Phi_j$ at $t=T$, 
and is expressed in the Euclidean Heisenberg picture as 
\begin{eqnarray}
W^{jk}_T&\equiv& \langle\Phi_j(T)|\Phi_k(0)\rangle 
=\langle \Phi_j|W_{\rm 3Q}(T)|\Phi_k\rangle
=\langle\Phi_j|e^{-\hat HT}|\Phi_k\rangle \nonumber\\
&=&\sum_{m=0}^\infty \sum_{n=0}^\infty \bar c_m^j c_n^k
\langle m|e^{-\hat HT}|n \rangle
=\sum_{n=0}^\infty \bar c_n^j c_n^k e^{-V_nT}.
\end{eqnarray}
We define the matrix $C$ and the diagonal matrix $\Lambda_T$ by 
$
C^{nk} =c_n^k, 
\quad \Lambda_T^{mn}=e^{-V_nT}\delta^{mn}, 
$
 and rewrite the above relation as 
\begin{eqnarray}
W_T=C^\dagger \Lambda_T C.
\label{WandLambda}
\end{eqnarray}
Note here that $C$ is not a unitary matrix, and hence this relation does not mean 
the simple diagonalization by the unitary transformation, which
connects the two matrices by the similarity transformation.

Since we are interested in the 3Q potential $V_{n} \ (n=0,1,2,..)$ 
in $\Lambda_T$ rather than 
the non-universal matrix $C$, we single out $V_{n}$ 
from the 3Q Wilson loop $W_T$ using the following prescription. 
From Eq.(\ref{WandLambda}), we obtain 
\begin{eqnarray}
W^{-1}_TW_{T+1}&=&\{C^\dagger \Lambda_T C\}^{-1} C^\dagger \Lambda_{T+1} C
\nonumber\\
&=&C^{-1}\Lambda_T^{-1}\Lambda_{T+1}C\nonumber\\
&=&C^{-1}{\rm diag}(e^{-V_0},e^{-V_1},..)C.
\end{eqnarray}
Now the r.h.s is expressed by the l.h.s 
with the similarity transformation.
Therefore, $e^{-V_n}$ can be obtained 
as the eigenvalues of the matrix $W_T^{-1}W_{T+1}$, 
i.e., the solutions of the secular equation, 
\begin{eqnarray}
{\rm det}\{W_T^{-1}W_{T+1}-tI\}
&=&{\rm det} \{\Lambda_T^{-1}\Lambda_{T+1}-tI\}\nonumber\\
&=&\prod_{n}
(e^{-V_n}-t)=0, 
\label{secular}
\end{eqnarray}
where $I$ denotes the unit matrix.
The largest eigenvalue corresponds to 
$e^{-V_0}$ and the $n$-th largest eigenvalue corresponds to $e^{-V_n}$.

In this way, the 3Q potential $V_n \ (n=0,1,2,..)$ can be obtained from 
the matrix  $W_T^{-1}W_{T+1}$.
In the practical calculation, we prepare $N$ independent sample operators 
$|\Phi_k \rangle \ (k=0,1,..,N-1)$. 
If one chooses appropriate operators $|\Phi_k \rangle$ 
so as not to include highly excited-state components, 
one can truncate the physical states as $|n \rangle \ (n=0,1,2,..,N-1)$. 
Then, $W_T$, $C$ and $\Lambda_T$ are reduced into $N\times N$ matrices, 
and the secular equation Eq.(\ref{secular}) becomes the $N$-th order equation. 

Now, we study the ground-state potential $V_{\rm 3Q}^{\rm g.s.}$ 
and the 1st excited-state potential 
$V_{\rm 3Q}^{\rm e.s.}$ for the spatially-fixed 3Q system 
in SU(3) lattice QCD with $\beta$=5.8 and $16^3 \times 32 $ at the quenched level. 
The lattice spacing is found to be $a \simeq$0.14fm, which 
reproduces the string tension $\sigma$=0.89GeV/fm in the Q-$\bar{\rm Q}$ potential.

In this paper, we concentrate ourselves on the ground state $|0 \rangle$ and 
the 1st excited state $| 1 \rangle$ in the spatially-fixed  3Q system. 
To extract $V_0$ and $V_1$, we need to prepare at least two independent operators $|\Phi_k\rangle (k=0,1)$, 
and construct the $2 \times 2$ matrix $W_T^{-1}W_{T+1}$ with them. 
Here, the sample operators $|\Phi_k\rangle$  
can be freely chosen, as long as they satisfy the two conditions: 
the linear independence and the smallness of the higher excited-state components $|n\rangle$ with $n \ge 2$, 
which leads to $|\Phi_k \rangle \simeq c_0^k|0\rangle+c_1^k|1\rangle$.

As the sample 3Q operators $|\Phi_k\rangle$, 
we adopt the properly smeared 3Q operators  
since the higher excited-state components are reduced in them~\cite{TSNM02}.
After some numerical check on the above two conditions, 
we adopt the 8th, 12th, 16th, 20th smeared 3Q operators as the candidates of the sample 3Q operators. 
(Owing to the intervals of  4 iterations, these smeared operators are clearly independent each other.
The $N_{\rm smr}$-th smeared operator with $N_{\rm smr}\ge 8$ has small higher excited-state components.)

\begin{figure}[h]
\begin{center}
\epsfile{file=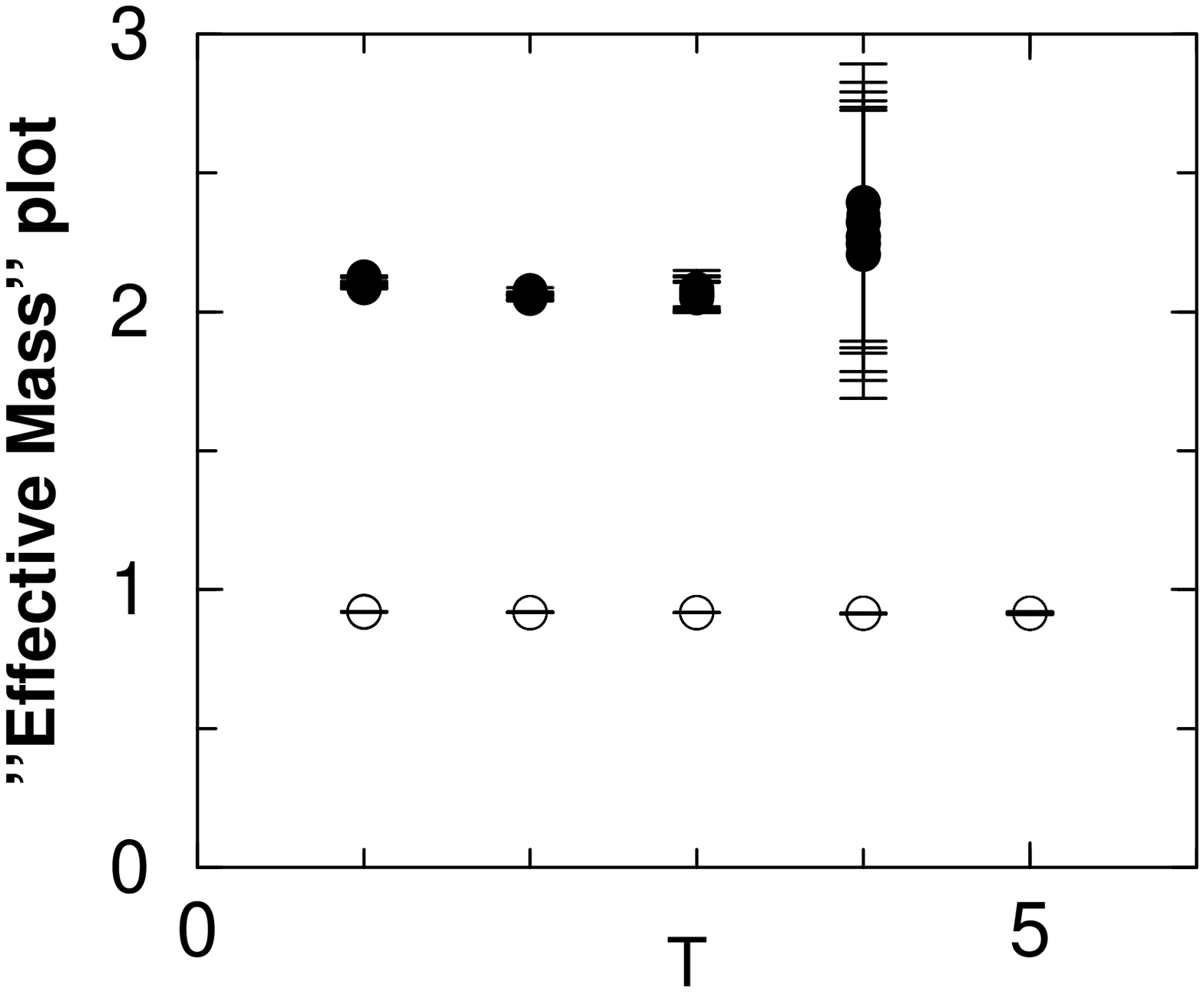,scale=0.4}
\caption{\label{EMplot}
An example of the effective-mass plot for $V_0(T)$ (open circles) and 
$V_1(T)$ (filled circles) 
obtained from the eigenvalues of $W_T^{-1}W_{T+1}$ at each $T$ 
for the spatially-fixed three quarks 
put on $(1,0,0)$, $(0,1,0)$ and $(0,0,1)$
in the lattice unit.
For both $V_0(T)$ and $V_1(T)$ at each $T$, we plot 6 data obtained from the 6 pairs of the operator combination.
}
\end{center}
\end{figure}

For each 
pair of these 4 operators, we calculate the generalized 3Q Wilson loop 
$W_T^{jk} \equiv \langle\Phi_j(T)|\Phi_k(0)\rangle$, and  evaluate $V_0$ and $V_1$ using Eq.(\ref{secular}). 
We plot in Fig.~\ref{EMplot} an example of the ``effective mass'' plot for 
$V_0(T)$ and $V_1(T)$ obtained from Eq.(\ref{secular}) at each $T$ 
as the function of the temporal separation $T$.
On the statistical error of the lattice data, we adopt the jackknife error estimate.

For both $V_0(T)$ and $V_1(T)$ at each $T$, we obtain  the 6 data obtained from the 6 pairs 
of the operator combination among 
the 8th, 12th, 16th, 20th smeared 3Q operators. As shown in Fig.\ref{EMplot},  
these 6 data almost coincide and the $T$-dependence of $V_0(T)$ and $V_1(T)$ is rather small in a certain region of $T$.
This indicates the smallness of the higher excited-state components $|n \rangle$ with $n \ge 2$ in the sample operators 
because such contaminations provide a $T$-dependence in $V_0(T)$ and $V_1(T)$ and make the stability lost.

For the accurate measurement, we select the best pair of two operators  providing the most stable effective-mass plot, 
which physically means the smallest contamination of the higher excited states in them.
With the selected two operators, we extract the ground-state potential $V_{\rm 3Q}^{\rm g.s.}$
and the 1st excited-state potential $V_{\rm 3Q}^{\rm e.s.}$
using the $\chi^2$ fit as $V_{\rm 3Q}^{\rm g.s.}=V_0(T)$ and $V_{\rm 3Q}^{\rm e.s.}=V_1(T)$ 
in the fit range of $T$, where the plateau is observed.
We perform the above procedure for each 3Q system.

\begin{figure}[h]
\begin{center}
\epsfile{file=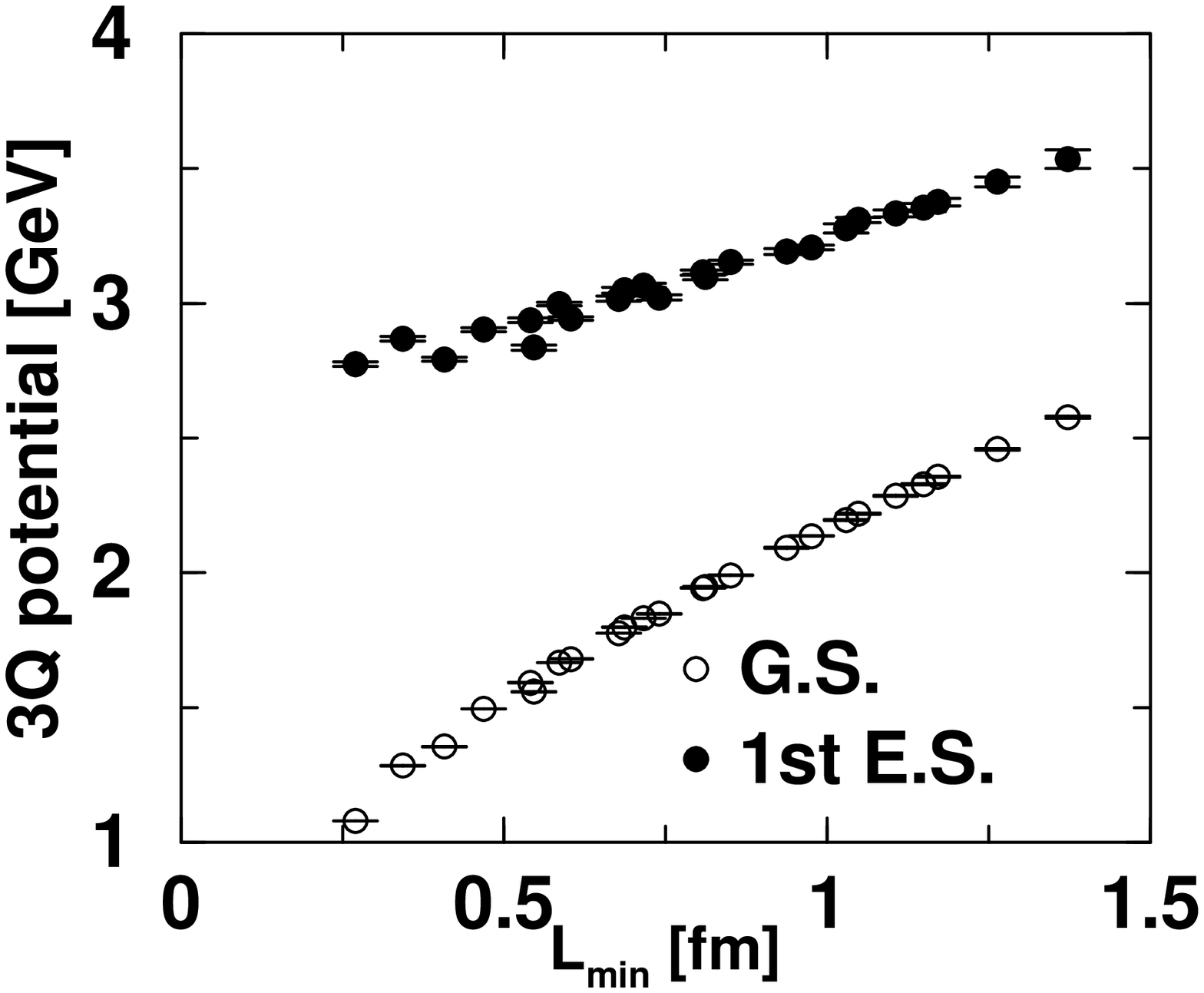,scale=0.4}
\end{center}
\caption{\label{3Qpot} The lattice QCD results of 
the ground-state 3Q potential $V^{\rm g.s.}_{\rm 3Q}$ (open circles) and the 1st 
excited-state 3Q potential $V^{\rm e.s.}_{\rm 3Q}$ (filled circles)  
plotted against 
$L_{\rm min}$, 
the minimal total length of flux-tubes linking the three quarks. 
}
\end{figure}

In Table~\ref{tab1}, we summarize the lattice QCD results for the 3Q potentials,
$V_{\rm 3Q}^{\rm g.s.}$ and $V_{\rm 3Q}^{\rm e.s.}$ for the 24 different patterns of the spatially-fixed 3Q system. 
In Fig.~\ref{3Qpot}, we plot the ground-state potential and 
the 1st excited-state potential as the function of $L_{\rm min}$ in the physical unit.

We attempt to compare $V_{\rm 3Q}^{\rm e.s.}$ with $V_{\rm Q\bar Q}^{\rm e.s.}$ or its 
linear combinations, 
since $V_{\rm 3Q}^{\rm e.s.}$ would exhibit a similarity to $V_{\rm Q\bar Q}^{\rm e.s.}$  
if the junction of the Y-type flux tube is a quasi-fixed edge.
However, there seems no simple relation between 
$V_{\rm 3Q}^{\rm e.s.}$ and $V_{\rm Q\bar Q}^{\rm e.s.}$,
which would reflect the complicated vibrational mode 
of the Y-type flux tube.
On the contrary, $V_{\rm 3Q}^{\rm e.s.}$ does not increase at the short distance, 
unlike the lowest gluonic-excitation mode in the Q-$\bar{\rm Q}$ system \cite{JKM98}.
Such a difference may indicate the quasi-free behavior of the junction of the Y-type flux tube.

\begin{table}[h]
\begin{center}
\newcommand{\m}{\hphantom{$-$}}
\newcommand{\cc}[1]{\multicolumn{1}{c}{#1}}
\renewcommand{\tabcolsep}{0.3pc} 
\renewcommand{\arraystretch}{1} 
\caption{\label{tab1}
The ground-state 3Q potential $V_{\rm 3Q}^{\rm g.s.}$
and the 1st excited-state 3Q potential $V_{\rm 3Q}^{\rm e.s.}$ in the lattice unit. 
The label $(l, m, n)$ denotes the 3Q system 
where the three quarks are put on $(la, 0, 0)$, $(0, ma, 0)$ and $(0, 0, na)$ 
in ${\bf R}^3$.
}
\small
\vspace{0.5cm}
\begin{tabular}{c c c c} \hline\hline
$(l, m, n)$ & $V_{\rm 3Q}^{\rm e.s.}$
& $V^{\rm g.s.}_{\rm 3Q}$ & $V^{\rm e.s.}_{\rm 3Q}-V_{\rm 3Q}^{\rm g.s.}$ \\
\hline
(0,1,1) &  1.9816( 95)  &  0.7711( 3)  &  1.2104\\
(0,1,2) &  1.9943( 72)  &  0.9682( 4)  &  1.0261\\
(0,1,3) &  2.0252( 92)  &  1.1134( 7)  &  0.9118\\
(0,2,2) &  2.0980( 80)  &  1.1377( 6)  &  0.9603\\
(0,2,3) &  2.1551( 87)  &  1.2686( 9)  &  0.8866\\
(0,3,3) &  2.2125(114)  &  1.3914(13)  &  0.8211\\
(1,1,1) &  2.0488( 90)  &  0.9176( 4)  &  1.1312\\
(1,1,2) &  2.0727( 75)  &  1.0686( 5)  &  1.0041\\
(1,1,3) &  2.1023( 73)  &  1.2004( 7)  &  0.9019\\
(1,1,4) &  2.1580( 93)  &  1.3201(10)  &  0.8380\\
(1,2,2) &  2.1405( 72)  &  1.1907( 7)  &  0.9498\\
(1,2,3) &  2.1899( 71)  &  1.3084( 9)  &  0.8815\\
(1,2,4) &  2.2516( 79)  &  1.4221(12)  &  0.8296\\
(1,3,4) &  2.2907( 91)  &  1.5260(15)  &  0.7647\\
(1,4,4) &  2.3807(138)  &  1.6322(20)  &  0.7485\\
(2,2,2) &  2.1776(111)  &  1.2844(10)  &  0.8932\\
(2,2,3) &  2.2242( 96)  &  1.3882(11)  &  0.8360\\
(2,2,4) &  2.2799( 98)  &  1.4952(15)  &  0.7847\\
(2,3,4) &  2.3637(100)  &  1.5853(18)  &  0.7784\\
(2,4,4) &  2.4108(137)  &  1.6836(23)  &  0.7271\\
(3,3,3) &  2.3408(168)  &  1.5680(19)  &  0.7728\\
(3,3,4) &  2.3958(151)  &  1.6635(22)  &  0.7323\\
(3,4,4) &  2.4645(177)  &  1.7565(30)  &  0.7081\\
(4,4,4) &  2.5245(340)  &  1.8408(42)  &  0.6837\\
\hline\hline
\end{tabular}\\[2pt]
\end{center}
\end{table}

As a remarkable fact, the lowest gluonic-excitation energy 
is found to be about 1GeV in the hadronic scale as $L_{\rm min} \simeq 0.5-1.5{\rm fm}$.
This is rather large in comparison with the low-lying excitation energy
of the quark origin.
Also for the Q-${\rm \bar Q}$ system, the gluonic-excitation energy seems rather large \cite{JKM98}.
Owing to the large gluonic-excitation energy, the gluonic excitation mode is invisible in 
the low-lying excitations of hadrons, which is a reason of the 
success of the simple quark model without gluonic modes.
Such a gluonic excitation would contribute significantly 
to the highly-excited baryons with the mass above 2GeV, and  
the lowest hybrid baryon, which is described as $qqqG$ in the valence picture, 
is expected to have a large mass of about 2GeV.
In any case, the present lattice QCD data would provide the useful input 
in the potential model calculation for the hybrid baryons.

We have studied the gluonic excitation in the three-quark (3Q) system 
using SU(3) lattice QCD with $\beta$=5.8 and $16^3 \times 32$ at the quenched level. 
For the 24 different patterns of the spatially-fixed 3Q system, 
we have measured the gluonic excited-state potential, which corresponds to 
the flux-tube vibrational energy in the flux-tube picture.
As a result, we have found that 
the lowest gluonic-excitation energy is about 1GeV
in the hadronic scale, which is rather large 
in comparison with the low-lying excitation energy of the quark origin.
This large gluonic-excitation energy leads to 
the absence of the gluonic mode in the low-lying hadrons 
and brings about the success of the quark model.

One of the author (H.S.) is supported in part 
by Grant for Scientific Research (No,12640274) from the Ministry of Education, Culture, Science 
and Technology, Japan. The lattice QCD calculations have been performed on 
HITACHI-SR8000 at KEK.

\end{document}